\documentclass[twocolumn,showpacs,preprintnumbers,amsmath,amssymb]{revtex4}
\usepackage{amsfonts,amsmath,amssymb,amstext,latexsym,graphicx}
\newcommand{\bequ}{\begin{equation}}
\newcommand{\eequ}{\end{equation}}
\newcommand{\barr}{\begin{array}}
\newcommand{\earr}{\end{array}}
\newcommand{\bea}{\begin {eqnarray}}
\newcommand{\eea}{\end {eqnarray}}
\newcommand{\lb}{\label}

\begin{document}
\def \N {\mathbb{N}}
\def \Z {\mathbb{Z}}
\def \R {\mathbb{R}}
\def \C {\mathbb{C}}
\let\la=\lambda
\def \La {\Lambda}
\def \ka {\kappa}
\def \vphi {\varphi}
\def \Zd {\Z ^d}
\title{The low-lying energy-momentum spectrum for the four-Fermi model on a lattice}
\author{Petrus H. R. dos Anjos}\email{petrus@if.sc.usp.br}
\affiliation{Departamento de F\'{\i}sica e Inform\'atica, IFSC-USP\\
C.P. 369, 13560-970 S\~ao Carlos SP, Brazil}
\author{Paulo A. Faria da Veiga}\email{veiga@icmc.usp.br}\affiliation{Departamento de Matem\'atica Aplicada e
Estat\'{\i}stica, ICMC-USP\\C.P. 668, 13560-970 S\~ao Carlos SP,
Brazil}
 \date{\today}
\begin{abstract}
We obtain the low-lying energy-momentum spectrum for the
imaginary-time lattice four-Fermi or Gross-Neveu model in $d+1$
space-time dimensions ($d=1,2,3$) and with $N$-component fermions.
Let $\kappa>0$ be the hopping parameter, $\lambda>0$  the
four-fermion coupling and $M>0$ denote the fermion mass; and take
$s\times s$ spin matrices, $s=2,4$. We work in the $\kappa\ll 1$
regime. Our analysis of the one- and the two-particle spectrum is
based on spectral representation for suitable two- and four-fermion
correlations. The one-particle energy-momentum spectrum is obtained
rigorously and is manifested by $sN/2$ isolated and identical
dispersion curves, and the mass of particles has asymptotic value
$-\ln\kappa$. The existence of two-particle bound states above or
below the two-particle band depends on whether Gaussian domination
does hold or does not, respectively. Two-particle bound states
emerge from solutions to a lattice Bethe-Salpeter equation, in a
ladder approximation. Within this approximation, the $sN(sN/2-1)/4$
identical bound states have ${\cal O}(\kappa^0)$ binding energies at
zero system momentum and their masses are all equal, with value
$\approx -2\ln\kappa$. Our results can be validated to the complete
model as the Bethe-Salpeter kernel exhibits good decay properties.
\end{abstract}
\pacs{11.10.Kk,11.10.St,02.70.Hm,71.10.Fd} \maketitle
We consider an imaginary-time lattice functional integral
formulation of the four-Fermi or Gross-Neveu (GN) model with
$N$-flavor component fermion fields in $d+1$ dimensions, $d=1,2,3$,
and a global ${\rm U}(N)$ flavor symmetry.

The interest in such a model
is because it is very rich and presents a non-trivial behavior. For
example, in $d=1$, the GN model \cite{GN} is  renormalizable and
asymptotically free in the ultraviolet ($N>1$); it also exhibits
dynamical symmetry breaking with mass generation and dimensional
transmutation, and asymptotic completeness has been verified up to
the two-particle threshold \cite{also}. Also,
the GN model is simple enough to be, in principle, accessible via
analytical approaches \cite{ana}. It has been employed to test
properties of more complex field theories. It is a purely fermionic
model and, since fermions are bounded operators, it is in a special
position to apply (non-)perturbative techniques to obtain rigorous
control of some correlations properties, going beyond the level of
simply few perturbation orders \cite{rigorous}. Moreover, the GN
model is important for the description of interesting physical
phenomena as superconductivity (see e.g. Ref. \cite{SC}).

Let $\kappa\geqslant 0$ be the hopping parameter, $M>0$ the fermion
bare mass, $\lambda>0$ the four fermion coupling and take $s \times
s$ spin matrices. We work in the $\kappa\ll 1$ regime and analyze
the one- and two-particle sectors of the energy-momentum spectrum.
We show that there are $sN/2$ particles manifested by isolated
dispersion curves $w(\vec p)\equiv w(\vec p,\kappa,M,\lambda)$, up
to near the three-particle threshold (upper mass gap property). From
the ${\rm U}(N)$ symmetry these curves are identical; the particle
masses $m\equiv m(\kappa)\equiv w(\vec p=\vec 0)$ are all equal and
with asymptotic value $-\ln\kappa$. Two-particle bound states near
the two-particle band are obtained from solutions to a lattice
Bethe-Salpeter (B-S) equation, in a ladder approximation. Their
existence above or below the two-particle band depends on whether
Gaussian domination does hold or not, respectively. Within the
ladder approximation, the $sN(sN/2-1)/4$ bound state masses are all
equal and of order $-2\ln\kappa$, their binding energies are ${\cal
O}(1)$. The methods applied here, were established in Refs.
\cite{Schor,relative,Transfer,CMP,baryons} for lattice spin and
field theoretical models.
Also, our method has the main ingredients to extend the bound state
results beyond the ladder approximation (see Refs.
\cite{Transfer,CMP}).

Precisely, our system action is given by
\begin{equation} \label{action}
\begin{array}{ll}
 \!\!\!\!S(\bar \psi, \psi)\! =\!\! &(\kappa/2)  \sum_{x,\epsilon} \bar
 \psi_{\alpha, a} (x) \,\, \Upsilon_{\alpha \beta} ^{\epsilon e^\mu} \,\, \psi_{\beta, a} (x+\epsilon
 e^\mu)\vspace{.1cm}\\ & - M \sum_x \bar
 \psi_{\alpha, a} (x)\,\,\psi_{\alpha, a} (x)\vspace{.1cm}\\
 &  - (\lambda/N)
 \sum_x [ \bar \psi_{\alpha, a} (x)\,\,\psi_{\alpha, a} (x)]^2,
 \end{array}
\end{equation}
with summation over repeated indices, and where $\bar
 \psi_{\alpha, a} (x)$, $\psi_{\beta, b} (y)$ are Grassmann
 variables, $x = (x^0 ,\vec x)$ $\equiv(x^0 ,..., x^{d})$ is a lattice site,
 $ x^0 \in\Z_{1/2}\equiv \{\pm 1/2, \pm 3/2, \ldots\}$, $\vec x \in \Z^d$, $\alpha, \beta \in  \{1,2,..,s \}$
are spin indices , $a,b \in \{ 1,2,...N\}$ are flavor indices,
$\epsilon = \pm 1$ and, for $\mu = 0,...,d$, the $e^\mu$ are unitary
vectors of $\Z^{d+1}$. Originally, instead of $M$, the lattice
formulation gives us a bare mass ${\cal M}(\bar m,\kappa)=(\bar m +
2\kappa)$ but, without loss of generality, given $\kappa$ we choose
the bare fermion mass $\bar m={\cal O}(1)\gg \ka$ such that ${\cal
M}(\bar m, \kappa)=M>0$. Also, $\Upsilon^{\pm e^\mu} = -I \pm
\varphi^\mu$, where $\varphi$ are $d+1$ hermitian traceless
anti-commuting matrices. For $d=2,3$, we take $s=4$ and
$\varphi^\mu\equiv\gamma^\mu$, $\mu=0,1,\ldots,d$, are $4 \times 4$
Euclidean Dirac matrices $\gamma^0 = {\scriptsize
\left(\begin{array}{cc} I_2 & 0 \\ 0 & -I_2
\end{array}\right)}$, $\gamma^j= {\scriptsize \left(\begin{array}{cc}  0 &
-i\sigma^j \\ i\sigma^j& 0
\end{array} \right)}$, and $\sigma^j, \, j=1,2,3$ are the hermitian
traceless anti-commuting Pauli matrices. In these cases, we have of
course $s=4$. For $d=1,2$, we also consider the choice
$\varphi^0=\sigma^3$, $\varphi^{j=1,...,d}=\sigma^j$ being Pauli
matrices, and $s=2$. From now on, we concentrate our analysis here
only on the $d=3$, $s=4$ case. The other cases can be reproduced
from this one.

We start in a finite-volume $\Lambda\subset \Z_0^{d+1}\equiv
\Z_{{1}/{2}} \otimes \Z^{d}$. The normalized expectation of a
function $F(\bar \psi,\psi)$ is
$$\langle F(\bar \psi,\psi)\rangle_{\Lambda} = \frac{1}{Z_\Lambda}\int_\Lambda
F(\bar \psi, \psi) e^{- S(\bar \psi, \psi)}\,d\bar \psi\,d\psi\,,$$
where $d\bar \psi\,d\psi=\prod_{ x,\,\alpha,\,a} d\bar \psi_{\alpha,
a} (x)\,d\psi_{\alpha, a} (x)$ and $Z_\Lambda = \int_\Lambda e^{-
S(\bar \psi, \psi)} d\bar \psi\,d\psi$. Using a polymer expansion
\cite{Seiler}, we show that correlations are given by convergent
expansions if $(\kappa/M)\ll 1$, and that the thermodynamic limit
$\Lambda\rightarrow\Z_0^{d+1}$ exists. The limiting correlation
functions (cf's) are denoted by $\langle \,\cdot\, \rangle$ and are
translation invariant and analytic functions jointly in $\kappa$ and
$\lambda$. As a consequence, the truncated cf's have exponential
tree decay with rate at least $-(1-\varepsilon)\ln [{\cal O}(1)
\kappa]$, where $\varepsilon \rightarrow 0$ as $\kappa \rightarrow
0$. Below, we work in the thermodynamic limit.

The choice of the shifted temporal lattice is to accommodate
two-sided equal-time limits of Fermi correlations in the continuum
limit. For $\kappa \geqslant 0$ the action of Eq. (\ref{action})
preserves positivity and, therefore, we can use a Feynman-Kac (F-K)
formula (see \cite{Seiler}) to construct an imaginary-time lattice
quantum field theory associated with this model. This construction
provides a positive metric Hilbert space ${\cal H}$, with inner
product $(F,G)_{\cal H}$, vacuum vector $\Omega$ and field operators
$\psi_\alpha^a(x), \bar \psi_\alpha^a(x)$. A straightforward
consequence of this construction is that the squared transfer matrix
$T_0^2=e^{-2H}$ is, in the infinite-volume limit, a well defined
self-adjoint contraction operator in the Hilbert space of physical
states ${\cal H}$ (assuming that $T_0$ does not have a $0$
eigenvalue). Thus, we can construct commuting self-adjoint
energy-momentum (E-M) operators $H \geqslant0$ and $\vec
P=(P^1,...,P^d)$ (with spectral points $\vec p\in{\bf T}^d\equiv
(-\pi,\pi]^d$), the generators of time and unitary space
translations $\vec T=e^{i\vec P.\vec x}$, respectively, and analyze
the joint spectrum of $T_0$ and $\vec P$.

For $F$ and $G$ with fixed time, and for the uppercase $\check .$
denoting an operator on ${\cal H}$, taking $x^0>0$, the F-K formula
is $(G,\check T_0^{x^0}\check T_1^{x^1}\check T_2^{x^2}\check
T_3^{x^3}F)_{{\cal H}} =\langle [T_0^{x^0}\vec T^{\vec x}F]\Theta
G\rangle$, where $T^{\vec x}=T_1^{x^1}... T_d^{x^d}$ and $\Theta$ is
an anti-linear operator which involves time reflection. The key idea
of our method to detect particles and bound states in the E-M
spectrum is to use the F-K formula to obtain spectral
representations for conveniently truncated two- and four-point cf's.
It is these spectral representations that allow us to relate the
complex momentum singularities of the Fourier transform of these
cf's to the E-M spectrum (see \cite{Transfer}). To obtain the
one-particle spectrum, we analyze the two-point cf ($x^0-y^0\not=0$)
\begin{equation}\label{2pt} \begin{array}{ll}G^{ab}_{\alpha\beta} (x,y)
=& \chi_{y^0\geqslant x^0}\left\langle \bar \psi_{\alpha a} (x)
\psi_{\beta b} (y)  \right\rangle\\ & - \chi_{y^0 < x^0}\left\langle
\psi_{\alpha a} (x) \bar \psi_{\beta b} (y)
\right\rangle^*,\end{array}
\end{equation}
where $\chi(x)$ is the Heaviside function. Using F-K and the
spectral representations of $T_0$ and $\vec P$, we obtain the
spectral representation for the two-point cf \bequ
G^{ab}_{\alpha\beta}(x,y)=\int_{-1}^{1}\int_{{\bf T}^{d}}
\lambda_0^{|x^0 - y^0|-1} e^{i \vec \lambda .(\vec x - \vec
y)}\,d\rho^{ab}_{\alpha\beta}(\lambda)\,,\lb{specG}\eequ where
$d\rho^{ab}_{\alpha\beta}(\lambda)=d(\bar \psi_{\beta a},{\cal
E}(\lambda) \bar \psi_{\alpha b} )_{\cal H}$ with ${\cal
E}(\lambda)={E_0}(\lambda_0)\prod_{j=1}^{d} {F_j}(\lambda_j)$, and
where ${E_0}(\lambda_0)$ (${F_j}(\lambda_j)$) are the spectral
families of $e^{-H}$ ($P^{j}$).

To obtain the two-particle spectrum, we analyze the partially
truncated four-point cf
\begin{equation}\label{4pt}
\begin{array}{ll} D^{ab}_{\alpha\beta}&\!\!\!(x_1, x_2,y_1,y_2) = \chi_{y^0_1 \leqslant x^0_1}\langle \bar \psi_{\alpha_1 a_1}
(x_1)\bar \psi_{\alpha_2 a_2} (x_2) \vspace{.15cm}\\&\times\:
 \psi_{\beta b_1} (y_1)\psi_{\beta_2 b_2} (y_2) \rangle- \chi_{y^0_1
>x^0_1}\langle \psi_{\alpha_1 a_1} (x_1) \vspace{.15cm}\\&
\times\:\psi_{\alpha_2 a_2} (x_2)\bar \psi_{\beta_1 b_1}
(y_1)\bar\psi_{\beta_2 b_2} (y_2)\rangle^*\,,
\end{array}
\end{equation}
where  $ab$ stands for $a_1a_2b_1b_2$, $\alpha\beta$ for
$\alpha_1\alpha_2\beta_1\beta_2$, and we work in the equal-time
representation $x^0_1=x^0_2$ and $y^0_1=y^0_2$. Using the lattice
relative coordinates $ \vec \xi = \vec x_2 - \vec x_1 ,\vec \eta =
\vec y_2 - \vec y_1, \tau = y_1 - x_2$ (see Ref.
\cite{relative,CMP}), we obtain, as for $G$, the spectral
representation ($\tau^0\not= 0$)
$$
D^{ab}_{\alpha\beta}(\vec \xi, \vec \eta, \tau)=
\int_{-1}^{1}\int_{{\bf T}^{d}} \lambda_0^{|\tau^0|-1} e^{i \vec
\lambda (\vec \xi + \vec \eta + \vec
\tau)}\,d\upsilon^{ab}_{\alpha\beta}(\lambda)\,,
$$where $d\upsilon^{ab}_{\alpha\beta}(\lambda)=
d(\phi_{\beta_2\beta_1}^{b_2b_1}(- \eta) , {\cal E}(\lambda)
\phi_{\alpha_1\alpha_2}^{a_1a_2}(\xi) )_{\cal H}$ and $\phi_{\alpha
\beta}^{a b}(x)= \bar \psi_{\alpha}^a(0) \bar
\psi_{\beta}^{b}(x)\Omega$.  Without spectral representations, the
analysis of the decay rates of cf's does {\em not} lead to particles
and bound states, specially when multiplicities are present with
narrow separations.

The cf's given in Eqs. (\ref{2pt}) and (\ref{4pt}) are matrix valued
operators on convenient spaces, and we can use symmetries at the
level of cf's to simplify them. We follow Ref. \cite{baryons} where
symmetries were analyzed. A symmetry operation $\mathbb{Y}$ on the
Grassmann algebra ${\cal G}$ of the fields $\psi$ and $\bar \psi$ is
an operation of ${\cal G}$ onto itself
 such that
$\mathbb{Y} \psi_{\alpha a}(x) = {\cal A}_{\alpha \beta} \tilde
\psi_{\beta b}({\cal Y}x)$, $\mathbb{Y} \bar \psi_{\alpha a}(x) =
\tilde{ \bar \psi}_{\beta b}({\cal Y}x){\cal B}_{\beta \alpha}$,
where ${\cal A}$ and ${\cal B}$ are $4\times4$ complex matrices and
${\cal Y}$ is a bijective linear map of $\Z_{0}^{d+1}$ onto itself.
The symbol $\sim$ stands for the introduction or the removal of a
bar. A symmetry of the model is a symmetry operation which leaves
invariant the action (\ref{action}) and the cf's (eventually apart
of a complex conjugation operation). Usual symmetries as discrete
spatial rotations, time {\it reversal} ${\cal T}$, charge
conjugation ${\cal C}$ and parity ${\cal P}$ can be implemented as
unitary (or anti-unitary for time reversal) operators in ${\cal H}$
are symmetries of the model. We also found a {\sl time reflection}
symmetry ${\textsc T}$, only at the level of cf's: ${\cal
Y}(x^0,\vec x) = (-x^0,\vec x)$
and ${\cal A}=i\left(\begin{array}{cc}0 & -\mathbb{I}_2 \\
\mathbb{I}_2 & 0
\end{array}\right)$, ${\cal B}={\cal A}^{-1}$.
By applying ${\cal C}$, the particle and the anti-particle cf's are
the same, and we only consider lower spin indexes, i.e.
$\alpha,\beta=3,4$ in the cf's of Eqs. (\ref{2pt}) and (\ref{4pt}).
Next, using ${\cal C}{\cal P}{\textsc T}$, we see that $G$ of Eq.
(\ref{2pt}) is a multiple of the identity. So, we drop the spin and
flavor indices (due to the ${\rm U}(N)$ symmetry!) from $G$.

Concerning the one-particle spectrum, we follow the analysis of
\cite{Schor}, for spin models. Using the hyperplane decoupling
method (\cite{S,baryons}), we see that $G(x,y)$ is bounded by
$|G(x,y)|< {\cal O}(1) \kappa^{|x-y|}$. Thus, taking $G=G_d+G_n$ as
a matrix operator on $\ell^2(\Z_0^{d+1})$, $G_d=G(x,x)\not= 0$, we
can define the inverse convolution matrix $\Gamma = G^{-1}$ through
a convergent Neumann series. Using the hyperplane decoupling method,
we can also show that the improved global bound $|\Gamma(x,y)\equiv
\Gamma(x-y)|<{\cal O}(1)\; \kappa^{3|x-y|}$ holds for $|x-y|>1$.
Also, by a generalized Paley-Wiener theorem, the Fourier transform
$\tilde \Gamma (p) = \sum_x \Gamma(x)e^{-ip.x}$ is analytic up to
close to the three-particle threshold $\approx -3\ln \kappa$. Hence,
from Eq. (\ref{specG}), the one-particle spectrum is determined by
the zeroes of $\tilde\Gamma(p)$. By the analytic implicit function
theorem and a Rouch\'e's theorem argument, there is a zero of
multiplicity $sN/2$ (due to spin/flavor) corresponding to isolated
particle dispersion curves
\begin{eqnarray}\label{disp} \omega(\vec p)&=& -\ln g_2\kappa- g_2 \kappa
\sum_{j=1,\ldots,d} \cos p^j + {\cal O}(\kappa^2) \,,
\end{eqnarray} where $g_2 = \langle \bar
\psi_{\alpha a}(x)\psi_{\alpha a}(x)\rangle |_{\kappa=0}$. The
particle masses are $m\!\!\equiv\!\! m(\kappa)\!\!\equiv\!\!
w(\vec{p}\!=\!\vec{0})$ $=-\ln g_2\kappa -dg_2 \kappa + {\cal
O}(\kappa^2)$. Above the one-particle spectrum there are
two-particle,... finite width free-particle bands which eventually
overlap for a large particle number.

To search for two-particle bound states, we use a lattice B-S
equation for $D$. In operator form, it reads
\begin{equation}\label{B-S}D = D_0 + DKD_0\,.\end{equation} In
terms of its operator kernel, $D_0$ is given by
\begin{equation}\label{4ptfree}\begin{array}{lll}
D^{ab}_{0,\alpha\beta} (x,y)& = & G^{a_1b_2}_{\alpha_1
\beta_2}(x_1,y_2)G^{a_2b_1}_{\alpha_2 \beta_1}(x_2,y_1)\vspace{.15cm}\\
&& - G^{a_1b_1}_{\alpha_1 \beta_1}(x_1,y_1)G^{a_2 b_2}_{\alpha_2
\beta_2}(x_2,y_2)\,,
\end{array}
\end{equation}
and is the Gaussianly evaluated four-point function with the
corrected propagator. Formally, from Eq. (\ref{B-S}), $K = D_0^{-1}
- D^{-1}$. It is easy to show that $D$, $D_0$ are bounded operators
in $\ell^2_a(A)$, the antisymmetric subspace of $\ell^2(A)$, $A=\{
(x,y) \in \Z_0^{d+1}| x^0 = y^0 \}^{4N^2}$. Next, by a Neumann
series argument, we show that $D_0^{-1}$, $D^{-1}$ and hence
$K^{-1}$ exist as bounded operators in $\ell^2_a(A)$. This uses
decay properties of $D_0$.

Computing $K$ to the leading $\kappa$ (ladder) approximation
$K\equiv L$, and using this to solve Eq. (\ref{B-S}) and to find the
singularities of D, we obtain the bound states.

We now show that $L$ is local and ${\cal O}(\kappa^0)$. For this, we
write $D=D^0+D^T$. Thus, $D^T$ is the connected (truncated)
four-point cf, and vanishes at $\kappa=0$ for non-coincident points
(see Ref. \cite{Transfer}), as seen by writing $D^T$ in terms of
source derivatives of the logarithm of the generating function.
Using a Taylor expansion at $\kappa=0$, we find
$D^{ab}_{\alpha\beta} (x,y) = -2g_4 P_{xy}
(1-S_{\alpha\beta}^{ab})\mathbb{I}_A$  $- 2 g_2^2 (1-P_{xy})
\mathbb{I}_A+ {\cal O}(\kappa)$ and $D^{ab}_{0,\alpha\beta} (x,y) )=
-2g_2^2 \mathbb{I}_A$ $+ {\cal O}(\kappa),$ where $P_{x y} =
\delta_{x_1 x_2} \delta_{y_1 y_2} \delta_{y_1 x_2}$, $S_{\alpha
\beta}^{ab} = \delta_{\alpha_1 \alpha_2} \delta_{\beta_1 \beta_2}
\delta_{\beta_1 \alpha_2}$ $\delta_{a_1 a_2} \delta_{b_1 b_2}
\delta_{b_1 a_2}$, $\mathbb{I}_A$ is the identity on $\ell^2_a(A)$
and $g_4 =\langle \bar \psi_{\alpha a} (x)$$\bar \psi_{\beta a} (x)
\psi_{\beta a} (x)\psi_{\alpha a} (x) \rangle|_{\kappa=0}$. With all
this, $L^{a b}_{\alpha\beta} (x ,y) = -[
({2g_2^2})^{-1}-({2g_4})^{-1}]P_{xy} [ \mathbb{I}_{\alpha
\beta}^{ab} - S_{\alpha\beta}^{ab} ] =  -\aleph P_{xy} [
\mathbb{I}_{\alpha \beta}^{ab} - S_{\alpha\beta}^{ab} ], $ where
$\mathbb{I}$ is the $4N^2 \times 4N^2 $ identity.

In the ladder approximation, the B-S equation reads  $D^{a
b}_{\alpha\beta} (x,y)=$ $D^{a b}_{0,\alpha\beta}(x,y)$ $+\sum_z
 D^{a c}_{\alpha \zeta}(x,z)$ $L^{c d}_{\zeta \epsilon}(z,w)$ $D^{d b}_{0,\epsilon
 \beta}(w,y)$. Changing to the coordinates $\xi$, $\eta$ and $\tau$, taking the Fourier transform in $
\tau$ only, and setting $\tilde
 f(\vec \xi, \vec\eta, k^0) = \sum_\tau f(\vec \xi, \vec\eta, \tau)
 e^{-ik^0\tau^0}$, at zero spatial momentum, we find
 $ \tilde D^{a b}_{\alpha\beta}(\vec \xi ,\vec \eta, k^0) =$ $ \tilde D^{a b}_{0,\alpha\beta} (\vec \xi ,\vec \eta, k^0)$
 $- \tilde D^{a c}_{0,\alpha \zeta} (\vec \xi ,\vec 0,
 k^0)$
 $\aleph [
\mathbb{I}_{\epsilon\zeta}^{dc}$ $- S_{\epsilon\zeta}^{dc} ]$
$\tilde D^{d b}_{\epsilon
 \beta} (\vec 0 ,\vec \eta, k^0).$ Therefore,
 we obtain (suppressing $k^0$)
 $\tilde D^{a b}_{\alpha\beta}(\vec \xi ,\vec \eta) =$ $\tilde D^{a b}_{0,\alpha\beta} (\vec \xi ,\vec
 \eta)$
$+ \tilde D^{a c}_{0,\alpha\zeta} (\vec \xi ,\vec 0)$ $
\{[\mathbb{I}+ \aleph [ \mathbb{I} - S ] \tilde D_{0} (\vec 0 ,\vec
0) ]^{-1}\}^{cd}_{\zeta,\epsilon}$ $\tilde D^{d
b}_{0,\epsilon\beta}(\vec 0 ,\vec \eta).$ From the analysis of $G$,
$\tilde D^{a b}_{0,\alpha\beta} (\vec \xi ,\vec \eta,
 k^0)$ is analytic in $[0,2m(\kappa))\cup(2m(\kappa)
 +W,3m)$,
 where $W=2[w(\vec{\pi})-w(\vec{0})]=$ $8
\kappa g_2d+{\cal O}(\ka^2)$ is the width of the two-particle band.
Hence, there is no bound state for $\aleph =0$. Bound states occur
only when $\det \{\mathbb{I}+ \aleph [ \mathbb{I} - S ]$ $\tilde
D_{0} (\vec 0 ,\vec 0, k) \}=0$, and then are associated with a zero
eigenvalue of $\mathbb{I}+ \aleph \left[ \mathbb{I} - S \right]
\tilde D_{0}(\vec 0 ,\vec 0, k^0)$. By a straightforward
calculation, $D_0(\vec 0, \vec 0, \tau) = -2 G^2(\vec 0, \tau)
(\mathbb{I}-S)$, then we just have to obtain the eigenvalues of $
\mathbb{I}-2\aleph R(k^0) (\mathbb{I}-S)$, where
$R(k^0)\equiv\widetilde G^2(\vec 0,k^0)=\int \,\langle \bar \psi_{1
a} (0) \psi_{1 a} (\tau) \rangle^2 \, e^{-ik^{0}\tau_{0}}\,
d\tau_{0}d\vec{\tau}$. There are $4N^2$ eigenvalues. The eigenvalue
$\lambda_1=1$ is $(2N^2+N)$-fold degenerated, has eigenvectors in
the symmetric subspace of $\R^{4N^2}$ and does not satisfy the bound
state condition. The eigenvalue $\lambda_2= 1-2\aleph R(k^0)$ is
$(2N^2-N)$-fold degenerate and has eigenvectors in the antisymmetric
subspace of $\R^{4N^2}$ \cite{comm}. These states may satisfy the
bound-state condition. For them, the condition is given by
$R(k^0)=(2\aleph)^{-1}$. We next use the analysis of $G$ to write
Eq. (\ref{specG}) as ($x^0\not=0$) $ \label{espectral}
 G(0,x)=\!\int_{0}^{\infty}\!\!
 \int_{{\bf T}^{d}}  e^{i\vec{p}\cdot\vec{x}-E|x_{0}|}
d\sigma_{\vec{p}}(E)d\vec{p}$. Also, for $w(\vec p)$ in Eq.
(\ref{disp}), we use the decomposition $ d\sigma_{\vec{p}}(E)
 =Z(\vec{p},\kappa)\delta(E-\omega(\vec{p}))dE+d\hat{\sigma}_
{\vec{p}}(E)$, with $Z(\vec{p},\kappa)= \frac {\partial
 \tilde{\Gamma}}{\partial\chi}(p^{0}=i\chi,\vec{p})|_{
 \chi=\omega(\vec{p})}$, and
 where $\tilde{\Gamma}(p)=[\tilde
G(p)]^{-1}$. Both, $d\sigma_{\vec{p}}(E)$ and
$d\hat{\sigma}_{\vec{p}}(E)$, are
 positive measures. $d\hat{\sigma}_{\vec{p}}(E)$ has support in
$(\breve{m}, \infty)$, $\breve{m}\approx -3\ln \kappa$ is a lower
bound for the onset of three-particle spectrum. The decomposition of
$d\sigma_{\vec{p}}(E)d\vec{p}$ gives a separation into one-particle
and the three- or more particle contributions (given by
$d\hat{\sigma}_ {\vec{p}}(E)$). Thus, $Z(\vec{p},\kappa)>0$.

Proceeding as in Ref. \cite{CMP,Transfer}, and holding only the
product of one-particle contributions in $G^2$ (associated with
$Z(\vec{p},\kappa)\delta(E-\omega(\vec{p}))dE$), we obtain
$$R(k^0)
\!=\! 2(2\pi)^{d+1}\!\!\int_{{\bf
T}^{d}}\frac{\sinh(2\omega(\vec{p}))
[Z(\vec{p},\kappa)]^2}{\cosh(2\omega(\vec{p}))-\cos k^0}d\vec{p}
\!+\! {\cal O}(\kappa)\,,$$ where $Z(\vec{p},\kappa)\!=\!(2\pi)^{-d}
g_2 \!+\!{\cal O}(\kappa)$. So, for $0\!<\!\Im(k^0)\!<\!2m$ (i.e.,
below the two-particle threshold),  $R(k^0)>0$. Hence, there are
{\sl no} bound states for $\aleph<0$. Similarly, for
$2m+W<\Im(k^0)<3m$, $R(k^0)<0$ and again no bound state exists for
$\aleph>0$. The only $k^0$ singularities of $\tilde
D(\vec\xi,\vec\eta,k^0)$, for $\Im (k^0) \in (0,2m) \cup (2m+W,3m)$,
are solutions of $ 2\aleph \sum_{\tau}
G^2(0,\tau)e^{2m\tau^{0}}e^{-\Delta \tau^{0}}=1$,
and we have set $ ik^0=-2m+\Delta$ (recall $m$ is the mass).

We next give an intuitive argument for the bound state formula based
on the behavior of $\langle \bar \psi_{1 a} (0) \psi_{1 a} (\tau)
\rangle$. A rigorous argument follows from using the convolution
form in momentum space of the above condition and Eq. (\ref{specG}).
Expanding $\langle \bar \psi_{1 a} (0) \psi_{1 a} (\tau) \rangle$ to
the leading order in $\kappa$, we obtain $\langle \bar \psi_{1 a}
(0) \psi_{1 a} (\tau) \rangle \approx \kappa
^{|\tau^0|+|\vec{\tau}|} g_2 ^{|\tau^0|+|\vec{\tau}|+1}$. This
follows from expanding the numerator of the two-point cf. For the
fermion integration to be nonzero, there must be a chain of
overlapping bonds connecting the two points. Holding the chain of
minimal length, the above result follows. A polymer expansion is
used to rigourously control all contributions. Using the value of
$m$, we obtain the bound state condition, to the leading order in
$\kappa$, as $2g_2 \aleph (1-e^{-\Delta})=1$.

Let $\Delta^{A}=\Delta$ ($\Delta^{R}=-\Delta  - W$) denote the
binding energy at zero system momentum for the {\sl attractive}
({\sl repulsive}) case. For the attractive case, we find $\Delta^A =
- \ln [1-({2\aleph g_2})^{-1}]$, where
$\aleph=[g_4-g_2^2]/(2g_2^2g_4)>0$ ({\sl Gaussian Subjugation}, as
the four-point function dominates the square of the two-point
function), giving a bound state {\sl below} the two-particle band.
For the repulsive case, $ \Delta^R =\ln [1-({2\aleph g_2})^{-1}]- W
$, where now $\aleph <0$ ({\sl Gaussian Domination}), and the bound
state is {\sl above} the band.

In closing, we point out that the low-lying E-M spectra of other GN
models in $d=1$ have been obtained (at least {\sl approximately}) by
other methods in the continuum massless case. For instance, the
Bethe-Yang ansatz was employed in the chiral invariant version of
the GN hamiltonian, as well as semiclassical approaches as in Ref.
\cite{ana}. Our results agree qualitatively with those. However,
besides the fact they are obtained in a precise mathematical setting
and include the upper gap property (absent in hamiltonian
formulations), and that our method can be easily exported to other
models, our results can be extended beyond the ladder approximation
using the methods e.g. of Refs. \cite{Transfer,CMP}. To do this, a
fast decay of the B-S kernel is crucial. Using the hyperplane
method, we get $ |K_{\alpha \beta}^{ab} (\vec \xi, \vec \eta, \tau)|
\leqslant {\cal O}(1) \kappa^{4|\tau^0|+\frac{1}{2}(\|\vec \xi +
\vec \eta + 2 \vec \tau \| + \| \vec \xi \| + \| \vec \eta \|)}|$.
Whether or not the degeneracy of the bound states is preserved in
the complete model is a good question. Finally, we remark that the
spectral pattern obtained here for the ${\rm U}(N)$ GN model is
similar to the one found for the ${\rm O}(2N)$ spin model in Ref.
\cite{Transfer}, except that the later has also a different bound
state of multiplicity $2N$.

We acknowledge support from FAPESP and CNPq, and thank Professor M.
O'Carroll for discussions.


\begin{thebibliography}{100}
\bibitem{comm} The general degeneracy is given in the introduction.
\end{thebibliography}
\end{document}